# Mechanism of $O_2$ influence on the decomposition process of the eco-friendly gas insulating medium $C_4F_7N/CO_2$

Fanchao Ye, Yitian Chu, Pascal Brault, Dunpin Hong, Shuangshuang Tian, Yi Li, Song Xiao, Xiaoxing Zhang

***Abstract*—The $C_4F_7N/CO_2/O_2$ gas mixture is the most promising eco-friendly gas insulation medium available. However, there are few studies on the mechanism of the influence of the buffer gas $O_2$ ratio and its role in the decomposition characteristics of $C_4F_7N/CO_2$. In this paper, based on the ReaxFF reaction molecular dynamics method and density functional theory, a simulation of the thermal decomposition process of the $C_4F_7N/CO_2$ mixture under different $O_2$ ratios was carried out at temperatures in the range 2000 – 3000 K. A constructed model of the $C_4F_7N/CO_2/O_2$ mixture reaction system was used that included the possible reaction paths, product distribution characteristics and their generation rates. The calculation results show that the thermal decomposition of $C_4F_7N/CO_2/O_2$ mainly generates species such as $CF_3$, $CF_2$, CF, F, $C_2F_5$, $C_2F_4$, $C_2F_2$, $C_3F_7$, $C_2F_2N$, $C_3F_4N$, CFN, CN, CO, O, and C. Among them, the two particles $CF_2$ and CN are the most abundant. The first decomposition time of $C_4F_7N$ is advanced by the addition of $O_2$, while the amount of $C_4F_7N$ decomposed and the generation of major decomposed particles decreases. The addition of 0%-4% of $O_2$ decreases the reaction rate of the main decomposition reaction in the reaction system. Quantum chemical calculations show that the dissociation process occurring from the combination of $C_4F_7N$ with O atom is more likely to occur compared to the direct dissociation process of $C_4F_7N$ molecules. The conclusions of this study provide a theoretical basis for the optimization of the application ratio of $C_4F_7N/CO_2/O_2$ and the diagnosis of its equipment operation and maintenance.**

***Index Terms*—$C_4F_7N/CO_2/O_2$; Decomposition mechani sm; DFT; MD simulation**

## I. Introduction

Sulphur hexafluoride ($SF_6$) is widely used in all kinds of gas-insulated electrical equipment thanks to its excellent insulation properties and arc extinguishing ability [1]. However, $SF_6$ gas is an extremely potent greenhouse gas, with a GWP (Global Warming Potential) 23,500 times that of $CO_2$ and an atmospheric lifetime of up to 3,200 years [2][3]. As a possible $SF_6$ gas replacement, $C_4F_7N/CO_2/O_2$ gas mixture, the most promising eco-friendly gas insulating medium, has already demonstrated its ability in gas-insulated equipment in power systems around the world [4][5][6][7].

Some results have been obtained from experimental studies on the insulation, arc extinguishing and decomposition characteristics of $C_4F_7N/CO_2/O_2$ gas mixtures. Zhang et al. carried out short-circuit current breaking tests on a 40.5 kV ceramic column circuit breaker filled with a $C_4F_7N/CO_2/O_2$ gas mixture, and found that increasing the $O_2$ content improved the breaking performance of the gas mixture by 10% [8]. Earlier experimental studies on the insulating properties of $C_4F_7N/CO_2/O_2$ gas mixtures with different $O_2$ contents, air pressures and electric fields were carried out by Gao, Tu and Zhang et al., who analyzed the influence of different factors on the breakdown voltage and partial discharge inception voltage of the gas mixtures. They found that a certain content of $O_2$ could effectively improve the insulating properties of $C_4F_7N/CO_2$ gas mixtures [9][10][11][12]. Yang et al. investigated the decomposition characteristics of $C_4F_7N/CO_2/O_2$ gas mixture under suspension discharge and found that the inhibition effect on $C_4F_7N$ decomposition was most obvious when the volume fraction of $O_2$ was 4% [13]. Ye et al. investigated the local superheated decomposition of a $C_4F_7N/CO_2/O_2$ gas mixture and found that the volume fraction of all products except $C_3F_6$ decreased to different degrees when 2% $O_2$ was added to the $C_4F_7N/CO_2$ gas mixture [14].

Macro-level experimental studies have shown that the addition of a certain amount of $O_2$ can effectively improve the insulation, arc extinguishing performance and chemical stability of $C_4F_7N/CO_2$ gas mixtures [4][5][8]. The effect of

This work was supported in part by the National Natural Science Foundation of China under Grant 52107145. (PI: Xiaoxing Zhang, corresponding author of this paper).

Fanchao Ye is with the Hubei Engineering Research Center for Safety Monitoring of New Energy and Power Grid Equipment, Hubei University of Technology, Wuhan, 430068 China (yefanchao@hbut.edu.cn).
Yitian Chu is with the Hubei Engineering Research Center for Safety Monitoring of New Energy and Power Grid Equipment, Hubei University of Technology, Wuhan, 430068 China 1500637903@qq.com).
Pascal Brault is with the GREMI Laboratory at the University of Orleans in France and the French National Centre for Scientific Research (pascal.brault@univ-orleans.fr).
Dunpin Hong is with the GREMI Laboratory at the University of Orleans in France and the French National Centre for Scientific Research (dunpin.hong@univ-orleans.fr).
Shuangshuang Tian is with the Hubei Engineering Research Center for Safety Monitoring of New Energy and Power Grid Equipment, Hubei University of Technology, Wuhan, 430068 China (tianss@hbut.edu.cn).
Yi Li is with the School of Electricity and Automation, Wuhan University,430068 China (liyi_whuee@163.com).
Song Xiao is with the School of Electricity and Automation, Wuhan University,430068 China （xiaosongxs@gmail.com）.
Xiaoxing Zhang is with the Hubei Engineering Research Center for Safety Monitoring of New Energy and Power Grid Equipment, Hubei University of Technology, Wuhan, 430068 China (xiaoxing.zhang@outlook.com).

$O_2$ addition on the decomposition characteristics of $C_4F_7N/CO_2/O_2$ is essentially due to the large differences in the microchemical processes of the breakdown of $C_4F_7N/CO_2/O_2$ under the effect of changes in the $O_2$ content and other influencing factors, which could not be revealed by experimental studies. Fu et al. revealed the mechanism of the influence of O atoms on the decomposition process of $C_4F_7N$ by calculating the thermodynamic parameters of different reaction paths [15]. However, they did not consider the influence of $O_2$ molecules and their content variations on the decomposition process of the $C_4F_7N/CO_2$ reaction system. For larger reaction systems consisting of molecules with complex structures, molecular dynamics methods are suitable for calculating their chemical reaction processes. Zhang et al. and Liu et al. investigated the distribution pattern of decomposed particles of $C_4F_7N/CO_2$ gas mixtures based on the combination of molecular dynamics (MD) and quantum chemistry of the ReaxFF reaction [16][17]. They did not take into account, however, the effect of the addition of $O_2$ as a buffer gas on the decomposition process of the whole reaction system.

In this paper, the molecular dynamics simulation of the ReaxFF reaction was carried out through the constructed thermal decomposition model of the $C_4F_7N/CO_2/O_2$ gas mixture, and the composition, distribution and generation rate of particles of thermal decomposition of the gas mixture with different $O_2$ content and different temperatures were obtained. Based on density functional theory and transition state theory, the thermodynamic parameters of the decomposition process were obtained under the conditions of $O_2$ participation in the reaction. They revealed the decomposition process of $C_4F_7N$ and the mechanism of the product generation with the addition of $O_2$.

## Ⅱ. Methods

### A. ReaxFF calculation method

The ReaxFF reactive force field expresses the interactions between atoms in a reactive system as a function of bond level on the basis of the bond order (BO), i.e., the interaction energy of each part of the molecule is expressed in the form of bonding poles. The energy expression of the system is as follows [18]:

$$E_{system} = E_{bond} + E_{over} + E_{angle} + E_{torsion} + E_{vdWaals} + E_{coulomb} + E_{specific} \quad (1)$$

where $E_{system}$ is the total system energy; $E_{bond}$ is the bond energy term between the atoms; $E_{angle}$ and $E_{torsion}$ represents the three-bond correction energy term and the four-body energy action term, respectively; $E_{over}$ is an over-coordination energy correction term; $E_{vdwaals}$ and $E_{coulomb}$ are van der Waals and Coulomb terms, respectively, which describe the electrostatic and dispersive effects in the interaction, and are non-bonding terms; $E_{specific}$ is usually used for energy terms for specific systems, such as lone pair electrons, conjugated electrons, hydrogen bonding and C2 correction terms.

A 260Å*260Å*260Å periodic simulation box was first constructed. The whole reaction system contained 500 molecules including 75 $C_4F_7N$ molecules (15%). The ReaxFF-MD simulations were carried out by varying the $O_2$ content (0%, 2%, 4%, 6%, 8%, and 10%, considering the stability of $C_4F_7N$, the oxygen content should not exceed 10%) at a temperature of 2600 K and varying the simulated temperatures (2000 K, 2200 K, 2400 K, 2600 K, 2800 K, and 3000 K) for a given $O_2$ content (6%) of the $C_4F_7N/CO_2/O_2$ mixture. The composition of reactant particles in the simulated system, the temperature settings and the system density are shown in Table I

Table I

Modelling system reactant particle composition

| NO. | Number of $C_4F_7N$ molecules | Percentage | Number of $CO_2$ molecules | Percentage | Number of $O_2$ molecules | Percentage | Temperature | Density |
|---|---|---|---|---|---|---|---|---|
| 1 | 75 | 15% | 425 | 85% | 0 | 0% | 2600 | 0.003148 |
| 2 | 75 | 15% | 415 | 83% | 10 | 2% | 2600 | 0.003137 |
| 3 | 75 | 15% | 405 | 81% | 20 | 4% | 2600 | 0.003125 |
| 4 | 75 | 15% | 395 | 79% | 30 | 6% | 2600 | 0.003114 |
| 5 | 75 | 15% | 385 | 77% | 40 | 8% | 2600 | 0.003103 |
| 6 | 75 | 15% | 375 | 75% | 50 | 10% | 2600 | 0.003091 |
| 7 | 75 | 15% | 395 | 79% | 30 | 6% | 2000 | 0.003114 |
| 8 | 75 | 15% | 395 | 79% | 30 | 6% | 2200 | 0.003114 |
| 9 | 75 | 15% | 395 | 79% | 30 | 6% | 2400 | 0.003114 |
| 10 | 75 | 15% | 395 | 79% | 30 | 6% | 2800 | 0.003114 |
| 11 | 75 | 15% | 395 | 79% | 30 | 6% | 3000 | 0.003114 |

Firstly, based on the NVE system (i.e. the values of the number of atoms, the size of the reaction system and the energy are fixed), the geometry of the system was optimised at 20K for 10 ps (closed system, no heat exchange with the outside) to establish a more reasonable initial structure of the reaction. The optimized reaction systems of 15%$C_4F_7N$/85%$CO_2$ and 15%$C_4F_7N$/79%$CO_2$/6%$O_2$ are shown in Fig 1. Subsequently, equilibrium calculations were carried out for the system at 1000 K for 10 ps based on the NVT system (with constant values for the number of atoms, the size of the reaction system and the temperature). Finally, ReaxFF-MD was carried out for the $C_4F_7N/CO_2/O_2$ gas mixture system at different $O_2$ contents (0%-10%) and temperature intervals of 2000K-3000K. Considering the computational resources as well as the simulation results, the total duration of the simulation is 500 ps with a sampling step of 0.25 fs. The NVT system was used for simulation and the Berendsen temperature control method was chosen for the temperature control procedure. All Reax FF-MD simulations in this paper are based on the ReaxFF module in the cross-scale computational chemistry platform AMS (Amsterdam Density Functional) [19]. The force-field files were optimised using Monte Carlo methods based on NiCH.ff developed by Mueller et al. in conjunction with the properties of the $C_4F_7N$ molecule [20].

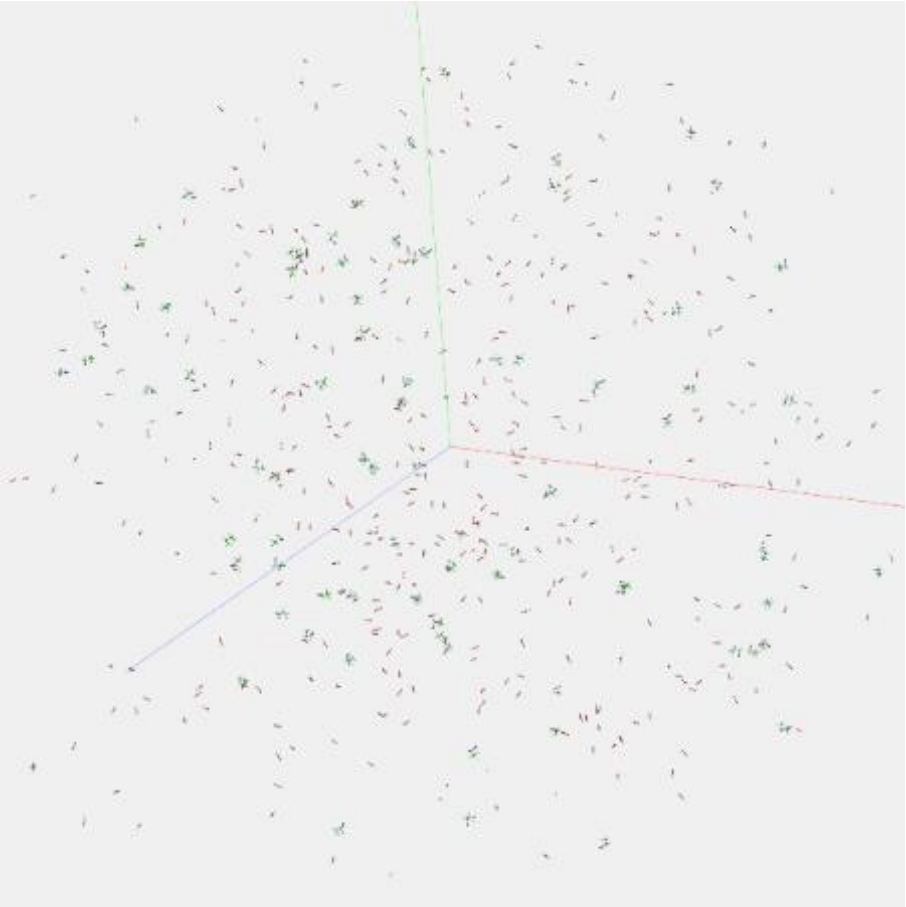

(a)

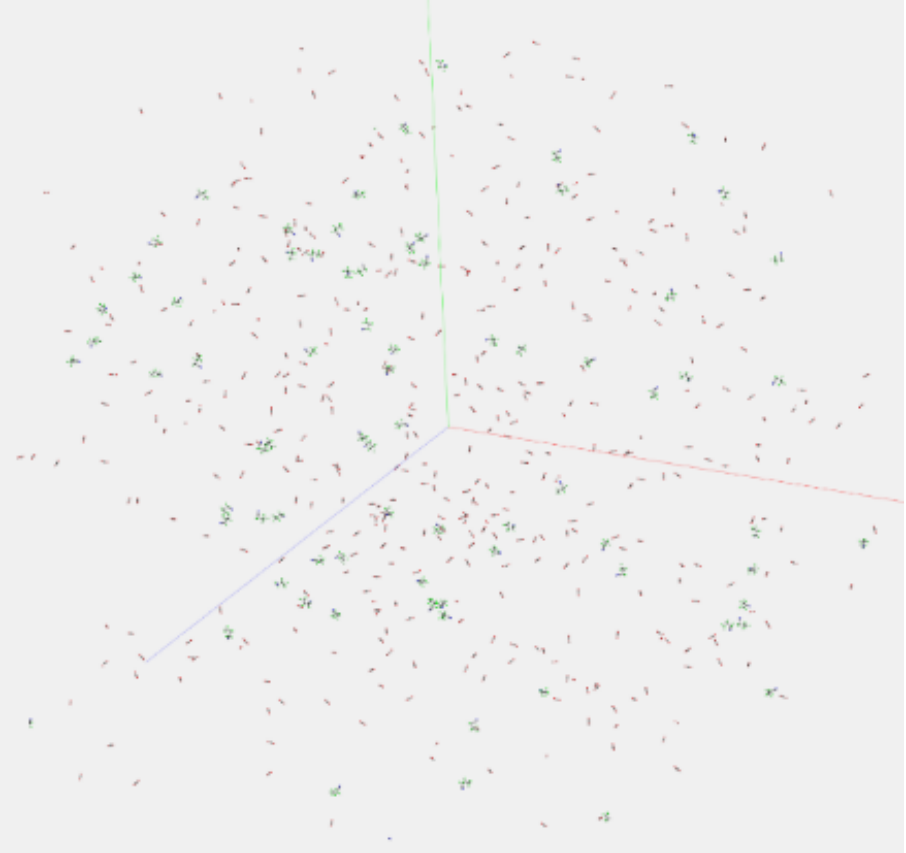

(b)

Fig. 1. Optimized reaction system model (a) 15%$C_4F_7N$-85%$CO_2$ (b) 15%$C_4F_7N$-79%$CO_2$-6%$O_2$

### B. Density functional theory

Density functional theory (DFT) and transition state (TS) theory based on quantum chemistry can provide theoretical support for calculating the microscopic parameters of the decomposition process of the decomposition process and revealing the decomposition of the $C_4F_7N$ discharge and the product generation mechanism under the $O_2$ participation in the reaction. In this paper, the decomposition mechanism of $C_4F_7N/CO_2/O_2$ was calculated based on DFT and TS. The Dmol$^3$ module of Materials Studio was used in order to obtain the energy and reaction enthalpies of the proposed reaction pathways. Geometry optimization and harmonic frequency calculation were based on the generalized gradient approximation and Perdew-Burke-Ernzerhof (GGA-PBE) methods [21]. Dual numerical plus polarization (DNP) was chosen as the basis group. The activation energy of the transition state reaction was obtained by searching the transition state of the proposed path using LST and QST methods [22].

## III. Results and discussion

### A. Kinetic analysis of the effect of $O_2$ on $C_4F_7N$ /$CO_2$/$O_2$ thermal decomposition particles

As the simulation temperature selected in this paper was 2000K-3000K, the temperature will not affect the reaction sequence or the reaction mechanism. However, in order to ensure completion of the chemical reaction process, a median temperature of 2600K was selected to carry out the kinetic analysis of the effect of $O_2$ on the thermal decomposition of $C_4F_7N/CO_2/O_2$ particles. The changes of $C_4F_7N$ and various types of decomposed particles in the $C_4F_7N/CO_2/O_2$ reaction system with different ratios of $O_2$ at 2600 K with simulation times are given in Fig. 2, and the number of each particle at the end of the simulated reaction is also given. It can be seen that a total of 18 particles, namely $CF_3$, $CF_2$, CF, F, $CF_3CF_2$ ($C_2F_5$), $CF_2CF_2$ ($C_2F_4$), $C_2F_2$, $(CF_3)_2CF$, $C_2F$, $CF_2CN$, $CF_3CFCN$ ($C_3F_4N$), CFN, CNO, CN, CO, O, C and $C_2$, were generated in the 15% $C_4F_7N$/85% $CO_2$ system at the end of the reaction. The number of decomposed particle species in the reaction system after the addition of $O_2$ depended on its ratio and was 14 (2% $O_2$), 15 (4% $O_2$), 16 (6% $O_2$), 13 (8% $O_2$) and 11 (10% $O_2$), respectively. The number of particle species in the reaction system basically shows a decreasing trend with the increase in $O_2$ content, indicating that the addition of $O_2$ can inhibit the generation of some particles to a certain extent. The main decomposition particles under different $O_2$ ratios are $CF_3$, $CF_2$, CF, F, CN and $C_2F_5$ (the number of particles is greater than 5). The production of O-containing products is relatively small, and $COF_2$ (carbonyloxyfluoride) is generated only when 6% $O_2$ and 8% $O_2$ are added, the number of $COF_2$ particles is 1 after 500 ps

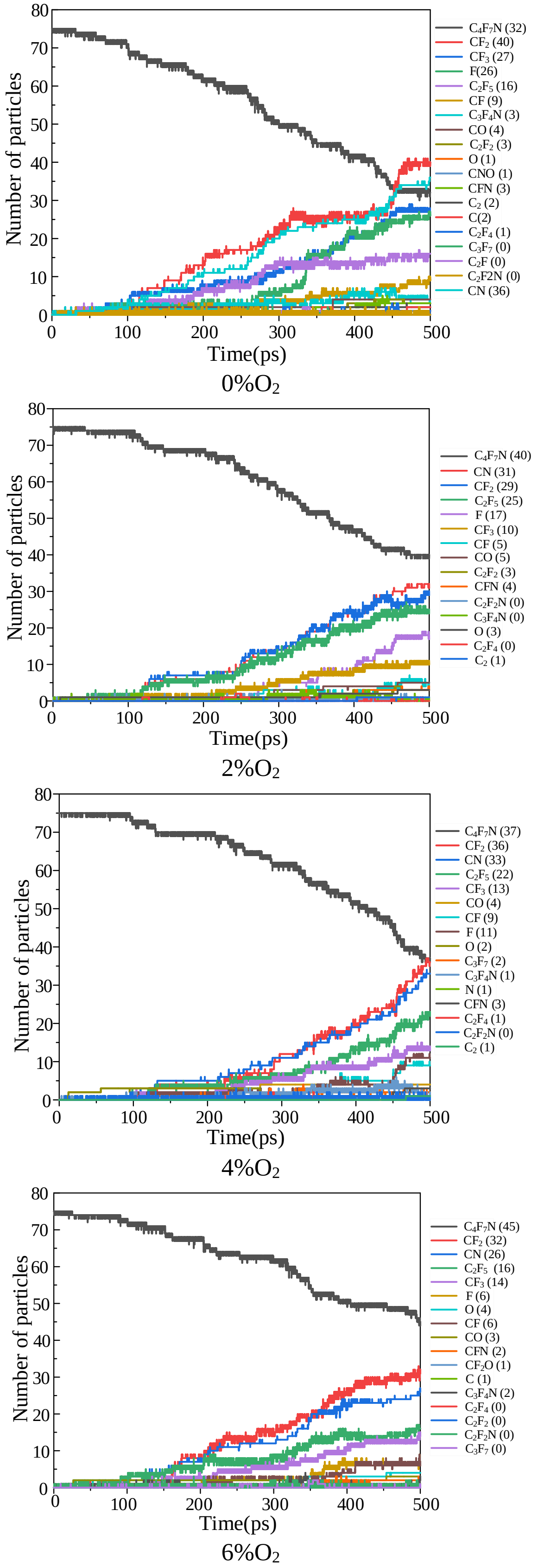

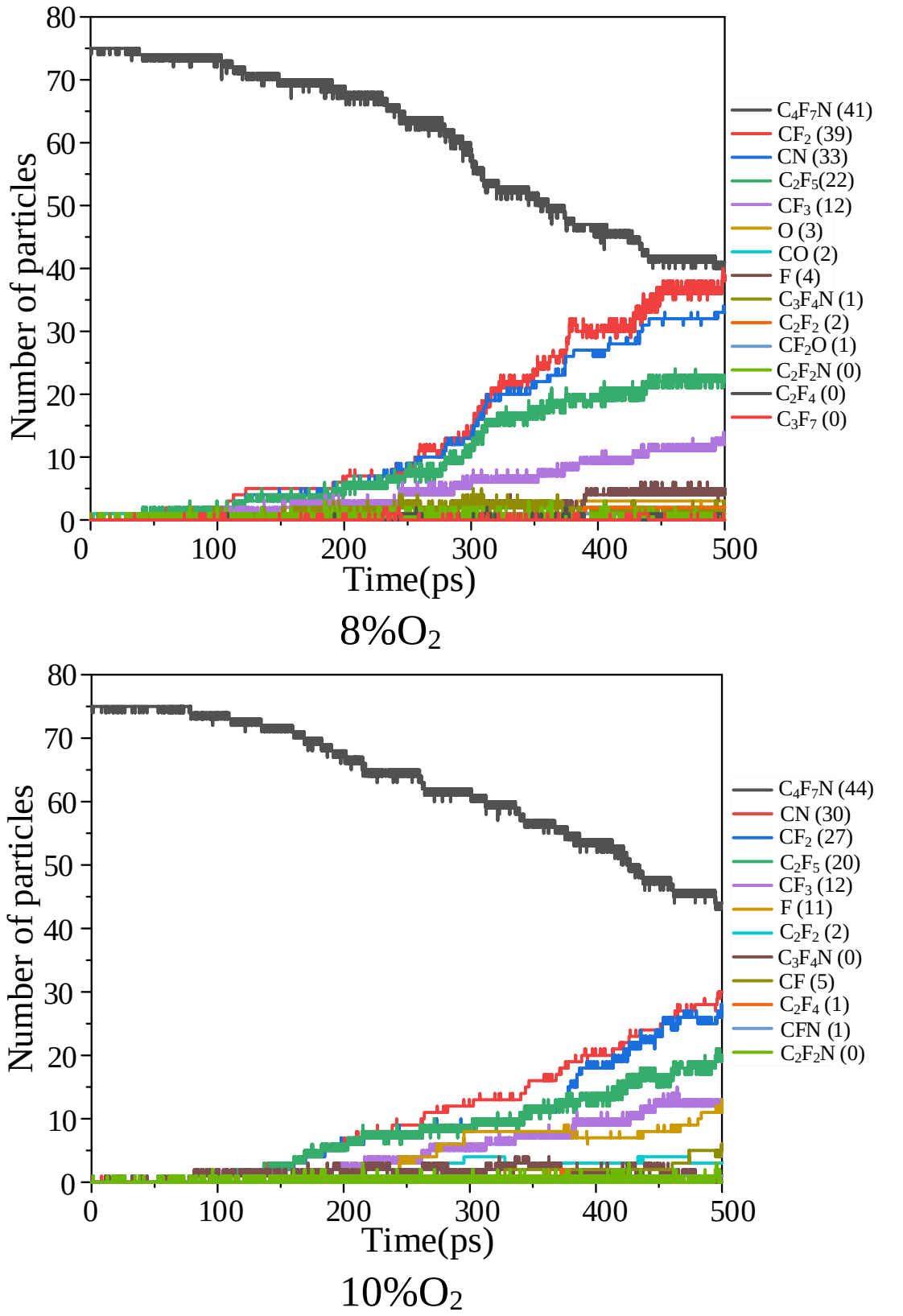

Fig. 2. Time evolution of $C_4F_7N/CO_2/O_2$ thermal decomposition particles with different $O_2$ contents at 2600k temperature. Numbers in brackets are the final values.

In the 15% $C_4F_7N$/85% $CO_2$ system, the decomposition of $C_4F_7N$ begins at 1.4 ps to form $CF_3$ and $CF_3CFCN$, and at 4.7 ps to form $C_3F_7$ and CN. The decomposition of $C_4F_7N$ to form $CF_3CF_2$ and $CF_2CN$ at a reaction time of 5.8 ps is not a simple one-time dissociation reaction: there is a transition state during the reaction. This reaction is more difficult than the dissociation reactions that occur at 1.4 ps and 4.7 ps. As the simulation time increases, $CF_3CFCN$ further decomposes to form $CF_3$, CF and CN, and $CF_3$ further decomposes to form $CF_2$ and F. Therefore, the maximum number of particles of $CF_3CFCN$ ($C_3F_4N$) at the end of the reaction is only 3 and decreases after the addition of $O_2$, suggesting that the addition of $O_2$ may accelerate the decomposition of $CF_3CFCN$, resulting in the generation of the predominantly several particles were CF3, $CF_2$, CF, F, CN and $C_2F_5$. When 2% $O_2$ was added to the $C_4F_7N/CO_2$ system, the decomposition of $C_4F_7N$ started to produce $CF_3CF_2$ and $CF_2CN$ at 0.75 ps and $CF_3$ and $CF_3CFCN$ at 1.1 ps, showing that the addition of $O_2$ can advance the initial dissociation time of $C_4F_7N$ compared with that of the case without $O_2$. Since the decomposition of $C_4F_7N$ to form $CF_3CF_2$ and $CF_2CN$ involves a transition state, it indicates that the addition of $O_2$ makes the decomposition process of $C_4F_7N$ more complicated from the early stage of the reaction.

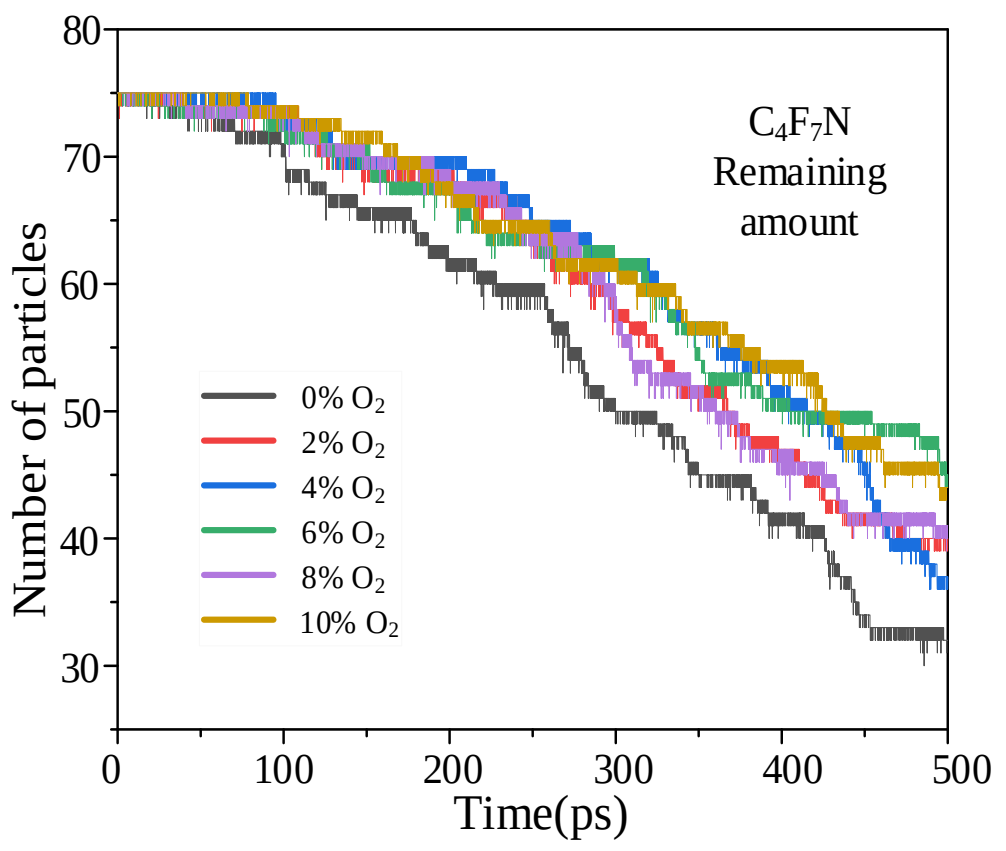

(a)

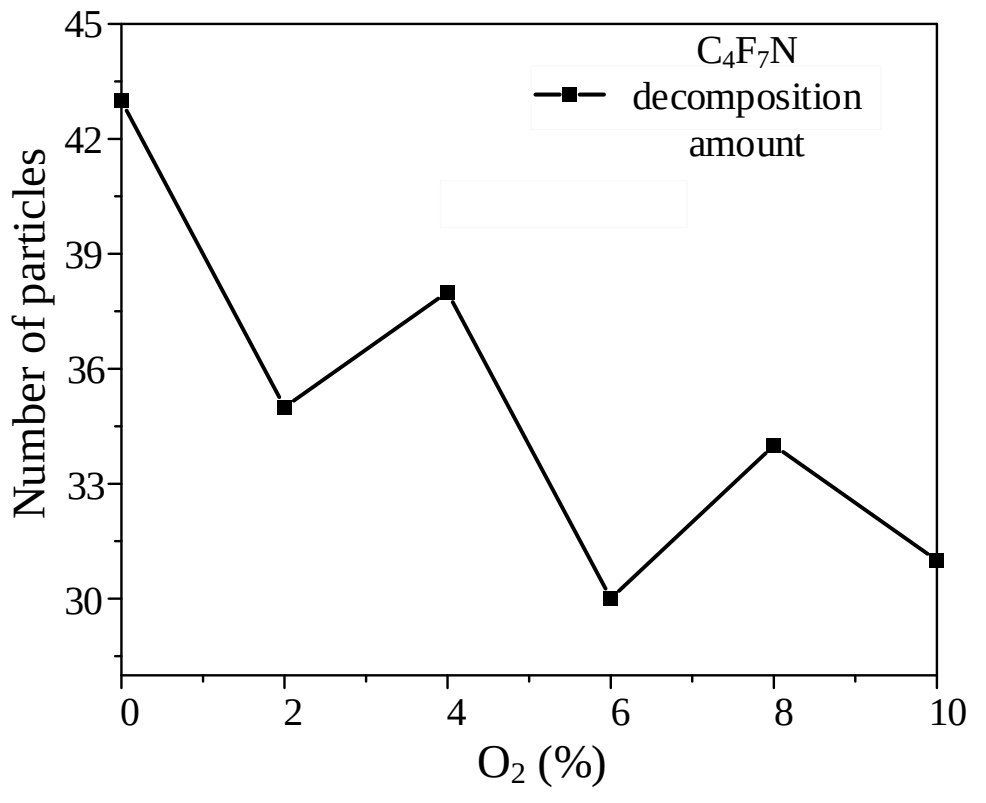

(b)

Fig. 3. Changes in $C_4F_7N$ particle number in reaction systems under different $O_2$ content (a) Trend with simulation time. (b) Trend with $O_2$ content at the end of the reaction

Fig 3(a) gives the variation of the number of $C_4F_7N$ particles in the reaction system with reaction time for different $O_2$ ratios. It can be seen that with the increase in simulation time, the decomposition of $C_4F_7N$ with different $O_2$ ratios gradually increases, so the number of $C_4F_7N$ particles in the reaction system decreases. It can be clearly seen that the decomposition rate (slope of the particle number versus time curve) accelerates at simulation times greater than 250 ps, especially in $O_2$-free conditions. Fig 3(b) gives the change in the decomposition amount of $C_4F_7N$ in the system at the end of the reaction with the $O_2$ content. It can be seen that after the addition of $O_2$, the decomposition amount of $C_4F_7N$ was clearly reduced and showed an overall decreasing trend with the increase in $O_2$ ratio, reaching a minimum at 6% $O_2$. Only 30 $C_4F_7N$ molecules were decomposed, which is 30.2% less compared with the decomposition of 43 $C_4F_7N$ molecules before the addition of $O_2$, indicating that the addition of a certain content of $O_2$ can effectively inhibit the decomposition of $C_4F_7N$.

In order to further investigate the influence of $O_2$ on the reaction rate of the system, the reaction rate constants k for some of the reaction paths involved in several major decomposing particles were calculated by the chemical reaction analysis algorithm ChemTraYzer 2 in the Advanced Work Flow module of the AMS software. The results are shown in Table II. As an important physical quantity in kinetic analysis, the rate constant k It can be seen that the reaction rates of paths 1-3 in Table II exhibit a decreasing trend at 0%-4% $O_2$ in the reaction system, and the reaction rate starts to increase for an $O_2$ ratio greater than 6%. At 10% $O_2$, there is an order of magnitude change in the reaction rate of path 3 compared to the other $O_2$ ratios. The reaction rate of path 4 decreases at 0%-8% $O_2$ in the reaction system and increases at 10% $O_2$. The addition of $O_2$ accelerates the reaction path 6. Overall, the addition of 0%-4% $O_2$ decreases the reaction rates of most of the decomposition reactions, but an $O_2$ ratio greater than 8% accelerates these reactions.

Table II.
DIFFERENT CONTENTS OF $O_2$ RATE CONSTANT (2600K)

| NO. | The reaction path | Rate constant, k ($cm^3 \cdot mol^{-1} \cdot s^{-1}$) | | | | | |
|---|---|---|---|---|---|---|---|
| | | 0% $O_2$ | 2% $O_2$ | 4% $O_2$ | 6% $O_2$ | 8% $O_2$ | 10% $O_2$ |
| 1 | $C_4F_7N \rightarrow CF_2CN+CF_3CF_2$ | 2.75E+09 | 2.04E+09 | 1.83E+09 | 1.92E+09 | 2.02E+09 | 2.83E+09 |
| 2 | $C_4F_7N \rightarrow CF_3+$ C $F_3CFCN$ | 2.03E+09 | 1.48E+09 | 1.47E+09 | 1.73E+09 | 1.71E+09 | 2.21E+09 |
| 3 | $CF_2CN \rightarrow CF_2+CN$ | 8.35E+11 | 7.63E+11 | 4.47E+11 | 4.66E+11 | 5.91E+11 | 6.15E+12 |
| 4 | $CF_2CF \rightarrow F+CFCF$ | 1.30E+12 | 1.03E+12 | 9.94E+11 | 5.76E+11 | 3.08E+11 | 1.33E+12 |
| 5 | $CF_3CF_2 \rightarrow CF_2+CF_3$ | 4.84E+09 | 3.25E+09 | 4.47E+09 | 9.18E+09 | 8.40E+08 | 5.45E+09 |
| 6 | $CO_2 \rightarrow CO+O$ | 9.47E+06 | 9.69E+06 | 1.51E+07 | 1.53E+07 | 1.05E+07 | - |

*B. Kinetic analysis of the effect of temperature on $C_4F_7N/CO_2/O_2$ thermal decomposition particles*

The variation in the number of $C_4F_7N$ particles in the reaction system with simulation time at different temperatures is given in Fig. 4(a). It can be seen that the decomposition rate of $C_4F_7N$ accelerates significantly with the increase in simulation temperature. Fig. 4(b) gives the variation in the amount of $C_4F_7N$ decomposition in the system at the end of the reaction with the simulation temperature. The number of particle species generated by the decomposition of $C_4F_7N$ at different

temperatures are 8 (2000K), 15 (2200K), 29 (2400K), 30 (2600K), 41 (2800K), and 48 (3000K), respectively, showing that at the simulation temperature in the range of 2200K-2400K, the decomposition rate of $C_4F_7N$ is faster, with little change between 2400K and 2600K, and that it continues to increase when it is greater than 2600K. This indicates that the decomposition of $C_4F_7N$ will reach a plateau when the simulated temperature increases to a certain interval, and that it is necessary to continue to increase the temperature in order to promote the decomposition of $C_4F_7N$.

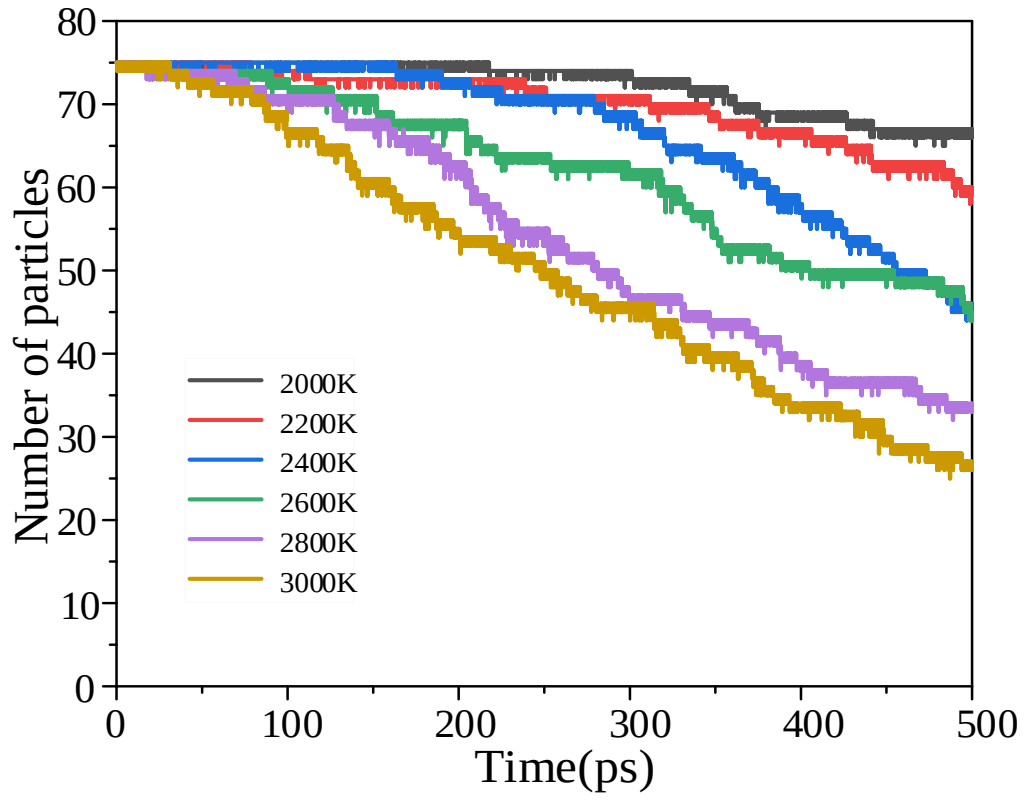

(a)

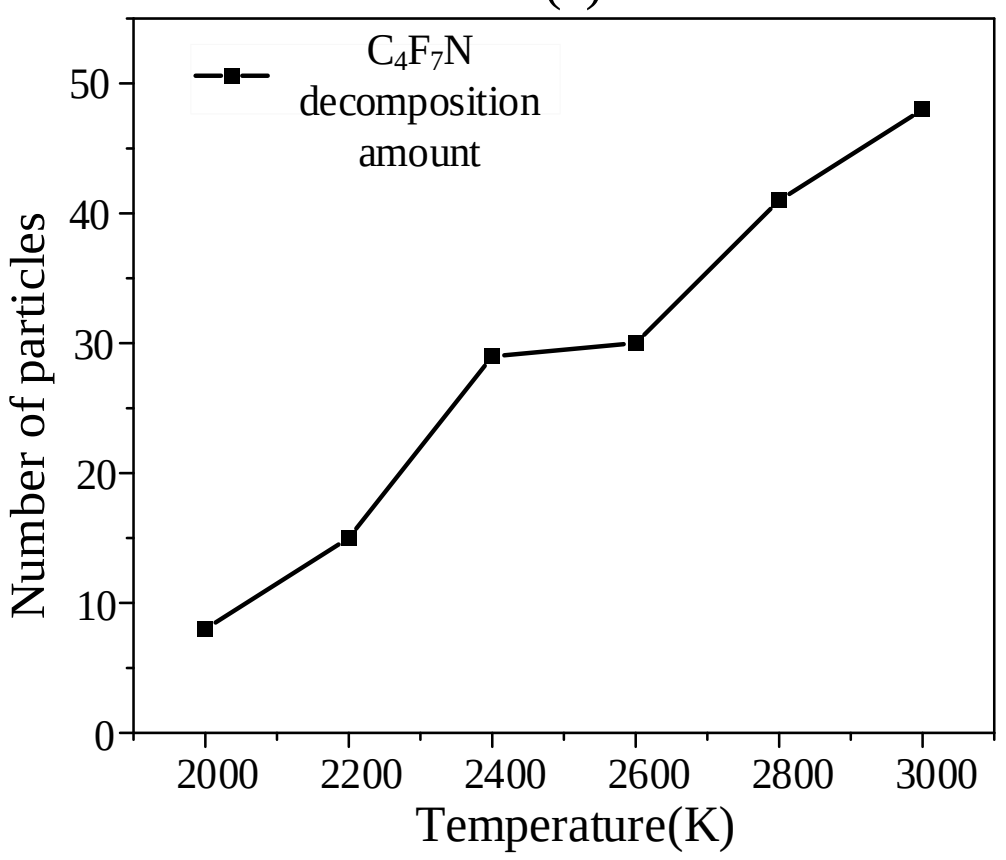

(b)

Fig.4. Changes in $C_4F_7N$ particle number under different temperature conditions (a) Trend with simulation time. (b) Trend with temperature at the end of the reaction

Table III shows the main thermal decomposition molecules of $C_4F_7N/CO_2/O_2$ containing 6% $O_2$ at different simulated temperatures. It can be seen that the two most abundant particles generated are $CF_2$ and CN, followed by $CF_3$ and F. The generation of $C_2F_5$ and CN shows a uniform increase with temperature and simulation time. It can be seen that the amount and rate of generation of $CF_3$ particles at the simulated temperature of 3000K is much larger than at other temperatures. $CF_3$ particles come from direct dissociation reactions of $C_4F_7N$, $CF_3CFCN$ and $C_2F_5$ particles on the one hand, and recombination of $CF_2$ and F particles on the other hand, and these dissociation and recombination reactions are intensified by an increase in the energy of the reaction system when the simulated temperature reaches 3000 K. The production rate of $CF_2$ is slow at the simulated temperatures of 2000K and 2200K. Its production rate increases substantially only when the simulation time is greater than 450 ps, and is accelerated significantly when the temperature is greater than 2200 K. The CF and F particles follow the same trend with temperature, and their production starts to increase rapidly at temperatures above 2400K.

Table III

GENERATION OF THE MAIN THERMAL DECOMPOSITION PARTICLES AT DIFFERENT SIMULATED TEMPERATURES (6%$O_2$)

| Temperature | Particle species | | | | | | | |
|---|---|---|---|---|---|---|---|---|
| (K) | $CF_3$ | $CF_2$ | CF | F | $C_2F_5$ | CN | O | CO |
| 2000 | 4 | 7 | 1 | 4 | 4 | 7 | 0 | 0 |
| 2200 | 1 | 13 | 3 | 2 | 14 | 14 | 0 | 0 |
| 2400 | 8 | 29 | 4 | 4 | 21 | 25 | 0 | 1 |
| 2600 | 14 | 32 | 6 | 6 | 16 | 26 | 4 | 3 |
| 2800 | 17 | 39 | 13 | 19 | 22 | 36 | 5 | 11 |
| 3000 | 29 | 44 | 16 | 27 | 17 | 45 | 1 | 4 |

Overall, the results of ReaxFF-MD simulations of the thermal decomposition process of the $C_4F_7N/CO_2/O_2$ mixture show that although the addition of $O_2$ to the $C_4F_7N/CO_2$ mixture decreases the initial decomposition time of $C_4F_7N$, it can effectively reduce the amount of decomposition and most of the particles generated, especially at an $O_2$ ratio of 6%, which results in the lowest amount of $C_4F_7N$ decomposition. Moreover, the addition of 0%-4% $O_2$ decreases the reaction rates of the major decomposition reactions in the reaction system, whereas $O_2$ levels greater than 8% accelerate these reactions. At simulated temperatures greater than 2600 K, the initial decomposition time of $C_4F_7N$ is significantly reduced and the rate of decomposition particle generation is accelerated. From the simulation results at different $O_2$ ratios and at different temperatures, it can be seen that the two most abundant particles generated are $CF_2$ and CN, followed by $CF_3$ and F.

### *C. Decomposition mechanism based on quantum chemical calculation*

From the above findings, it can be seen that the addition of $O_2$ as a background gas to the $C_4F_7N/CO_2$ mixture will have a significant effect on its decomposition particle distribution pattern and generation rate, so the dissociation of $C_4F_7N$ under the conditions of O particles participating in the reaction and the recombination process of each particle after dissociation and its thermodynamic parameters are of important significance. Density functional theory and transition state theory based on quantum chemistry will provide energy barrier heights as well as the chemical pathways under the conditions of $O_2$ addition, and further reveal the decomposition and product generation mechanism of $C_4F_7N$ with $O_2$ addition. To reveal the reaction mechanism of the $C_4F_7N/CO_2$ mixture with $O_2$ addition, 13 possible decomposition paths and 10 possible generation paths of O-containing products were constructed based on the structural characteristics of the $C_4F_7N$ molecule. The $C_4F_7N$ dissociation pathway with O addition is shown in Fig 5 (the molecules involved in the pathway are optimized configurations), and the enthalpies and activation energies under different pathways are shown in Table IV.

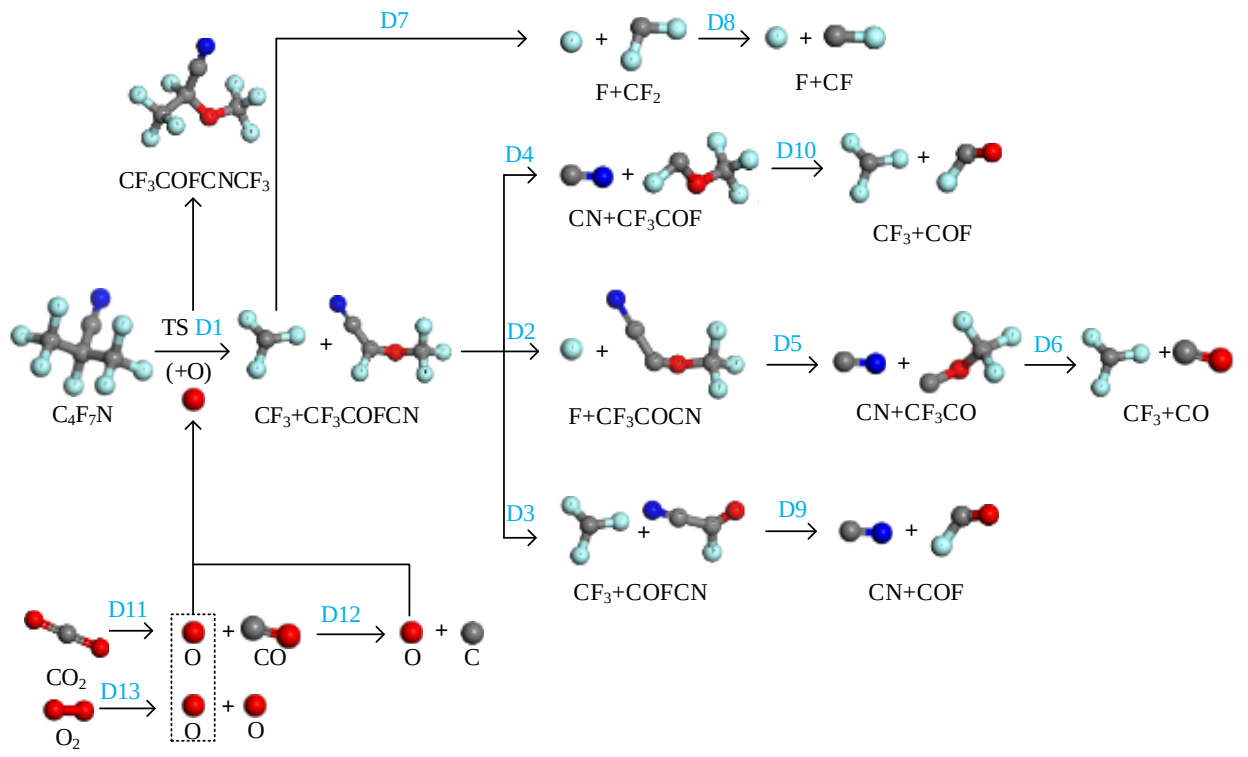

Fig. 5. $C_4F_7N$ dissociation pathways under O-atom reactions

According to the results of the molecular dynamics simulation of the reaction above, the decomposition process of $C_4F_7N$ becomes more complicated with the addition of $O_2$. In particular, the initial dissociation process is not a simple one-time dissociation process, but goes through a transition state. Firstly $O_2$ may react with $C_4F_7N$ in the form of O particles. The formation process and enthalpy of O particles are given in Fig 5 and Table IV, respectively, which show that O particles originate from the dissociation of $CO_2$ and $O_2$ in the gas mixture. The dissociation of $CO_2$ to generate O particles requires the absorption of 153.07 kcal/mol of energy, which is more likely to occur compared to the dissociation of $O_2$ to generate O particles, which requires the absorption of 182.29 kcal/mol energy. As the difference is not large, both are the main source of O particle generation. However, the further generation of C and O particles from CO requires the absorption of 514.60 kcal/mol of energy and occurs only at higher temperatures.

Table IV

THE ENTHALPY AND ACTIVATION ENERGY OF THE MAIN DISSOCIATION PATHWAY OF $C_4F_7N$ WITH O ADDITION

| No . | The reaction path | $\Delta_rH$ (kcal/mol) | ΔH (kcal/mol) |
|---|---|---|---|
| D 1 | $C_4F_7N$+O→$CF_3$+$CF_3COFCN$ | -106.52 | 47.51 |
| D 2 | $CF_3COFCN$ →$CF_3COCN$+F | 57.17 | |
| D 3 | $CF_3COFCN$ →$CF_3$+COFCN | 7.97 | - |
| D 4 | $CF_3COFCN$→$CF_3COF$+CN | 20.54 | - |
| D 5 | $CF_3COCN$ →C $F_3CO$+CN | 211.86 | |
| D 6 | $CF_3CO$ →$CF_3$+CO | -41.36 | |
| D 7 | $CF_3$→$CF_2$+F | 92.96 | |
| D 8 | $CF_2$→CF+F | 131.27 | |
| D 9 | COFCN →COF+CN | 134.38 | - |
| D10 | $CF_3COF$→$CF_3$+COF | 85.28 | - |
| D11 | $CO_2$→CO+O | 153.07 | |
| D12 | CO→C+O | 514.60 | |
| D13 | $O_2$→O+O | 182.29 | |

Path 1 in Table IV gives the initial dissociation process of $C_4F_7N$ when O particles are involved in the reaction. The reaction produces $CF_3$ and $CF_3COFCN$ particles. This reaction needs to undergo the transition state structure of $CF_3COFCNCF_3$, with the enthalpy of the reaction of -106.52 kcal/mol, and the need to overcome the energy barrier of 47.51 kcal/mol for this reaction to proceed properly. It has been shown that the dissociation of $C_4F_7N$ molecules to form $CF_3$ and $CF_3CFCN$ particles requires the absorption of 73.14 kcal/mol of energy [23]. The activation energy of path 1 is much lower than 73.14 kcal/mol, indicating that the participation of O particles in the reaction makes the bond breaking process of $C_4F_7N$ easier. Therefore, $O_2$ has a negative effect on the stability of $C_4F_7N$.

The $CF_3$ and $CF_3COFCN$ particles dissociate further, and the three primary dissociation paths (D2, D3 and D4) and their enthalpies for $CF_3COFCN$ are given in Table 4. The C=O bond breaking in $CF_3COFCN$ in path D3 to form $CF_3$ and COFCN requires an energy absorption of 7.97 kcal/mol, which is the lowest among the three paths, indicating that the C=O bond formed between the O atom in the centre of the $CF_3COFCN$ molecule and the $CF_3$ moiety has the lowest bonding energy. In path 2, dissociation of the $CF_3COFCN$ molecule requires absorption of the highest amount of energy (57.17 kcal/mol) among the three primary dissociation paths in order to dissociate the $CF_3COCN$ and F particles .

The particles such as $CF_3COCN$, $CF_3COF$ and COFCN generated by the primary dissociation of the $CF_3COFCN$ molecule are still macromolecules, which will be further dissociated under other molecules It can be seen that the energy required to further dissociate $CF_3COCN$ molecules to generate $CF_3CO$ and CN particles is higher (211.86 kcal/mol), indicating that the C-atoms of $CF_3COCN$ molecules indicating that the C-C bond connecting the CN group in the CF3COCN molecule is more stable. However, further dissociation of $CF_3CO$ to form $CF_3$ and CO particles can occur spontaneously and releases 41.36 kcal/mol of energy. The breaking of C=O in the $CF_3CO$ molecule to form $CF_3C$ and O particles is more difficult and requires the absorption of 209.4741.36 kcal/mol of energy. Therefore, $CF_3COFCN$ is more likely to decompose under sustained energy to form particles such as $CF_3$, CF and CN, which is in good agreement with the evolutionary pattern of the particles in the results of the molecular dynamics simulations above.

The possible formation pathways of the main O-containing products are given in Fig 6 (the molecules involved in the pathways are optimized configurations), and the enthalpies for the different pathways are shown in Table V. It can be seen that all O-containing products are generated by processes that can be spontaneous and exothermic (negative enthalpy), which may prevent the particles generated by the decomposition of $C_4F_7N$ from recombining to form $C_4F_7N$ or other stable by-products. The process of CO generation from C and O in pathway R10 releases the most energy (-236.4 kcal/mol), which explains why the addition of $O_2$ can inhibit the generation of carbon monomers from $C_4F_7N$ mixtures during the discharge process, as reported in the literature [4].

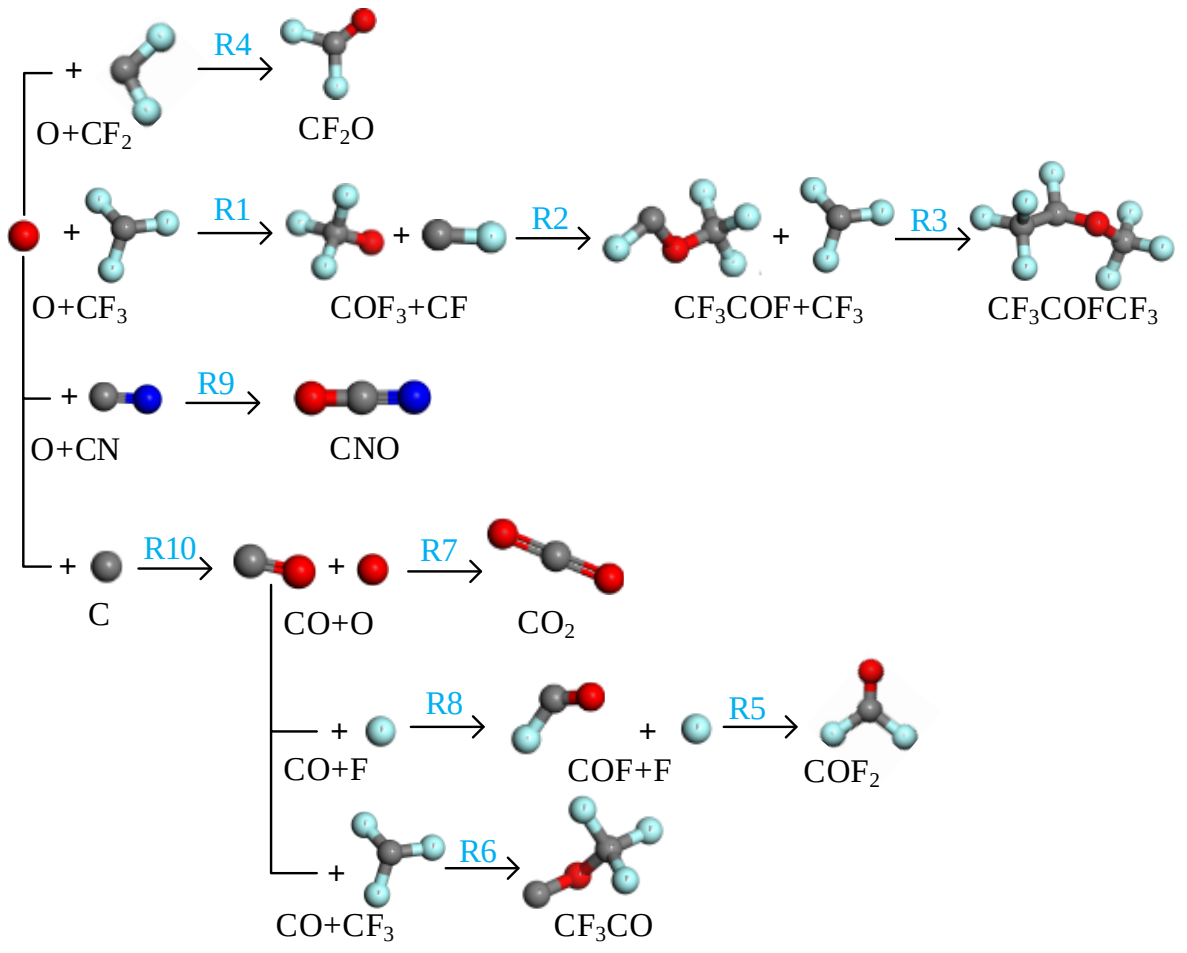

Fig. 6. Generation path of O-containing products

Table Ⅴ

MAIN O-CONTAINING PRODUCT FORMATION PATHS AND THEIR ENTHALPY VALUES

| No. | The reaction path | $\Delta_r H$ (kcal/mol) | $\Delta H$ (kcal/mol) |
|---|---|---|---|
| R1 | $CF_3+O \rightarrow COF_3$ | -104.05 | - |
| R2 | $COF_3+CF \rightarrow C_2F_4O$ | -72.59 | - |
| R3 | $C_2F_4O+CF_3 \rightarrow C_3F_7O$ | -161.13 | - |
| R4 | $CF_2+O \rightarrow COF_2$ | -108.72 | - |
| R5 | $COF+F \rightarrow COF_2$ | -120.89 | - |
| R6 | $CF_3+CO \rightarrow CF_3CO$ | -41.36 | - |
| R7 | $CO+O \rightarrow CO_2$ | -141.99 | - |
| R8 | $CO+F \rightarrow COF$ | -85.33 | - |
| R9 | $CN+O \rightarrow CNO$ | -177.12 | - |
| R10 | $C+O \rightarrow CO$ | -236.4 | - |

In summary, the energy released under superheated thermal failure conditions causes the simultaneous decomposition of $C_4F_7N$ and $O_2$ in the gas mixture to produce particles such as $CF_3$, $C_3F_7$, CN, $CF_2$, CF, F, $CF_3CF$, and O, which interact with each other to produce gas by-products. The addition of $O_2$ complicates the dissociation process of $C_4F_7N$, and the dissociation process of the product generated by the combination of $C_4F_7N$ and O occurs readily compared to the one-time dissociation process of the $C_4F_7N$ molecule. This suggests that the addition of $O_2$ will have an effect on the formation process of the by-products of the decomposition of $C_4F_7N$, and that the spontaneous reaction of the O particles with the decomposed particles of $C_4F_7N$ inhibits the generation of some of the by-products to a certain extent. Therefore, the addition of a certain amount of $O_2$ can reduce the generation of $C_4F_7N$ decomposition by-products, but as the spontaneous reaction between O particles and decomposition particles (e.g., fluorocarbons) generates $COF_2$ (strong toxicity), the generation of some of the by-products shows a tendency to increase when the $O_2$ ratio is higher.

## Ⅳ. CONCLUSION

In this paper, based on ReaxFF reaction molecular dynamics simulations and quantum chemical calculations, the mechanism of the influence of the buffer gas $O_2$ on the decomposition process of the environmentally friendly gas insulating medium $C_4F_7N/CO_2$ has been identified. The main conclusions are as follows:

(1) The ReaxFF-MD simulation results show that the thermal decomposition of the $C_4F_7N/CO_2/O_2$ gas mixture mainly generates particles such as $CF_3$, $CF_2$, CF, F, $C_2F_5$, $C_2F_4$, $C_2F_2$, $C_3F_7$, $C_2F_2N$, $C_3F_4N$, CFN, CN, CO, O, and C. The two particles generated in the highest amounts are $CF_2$ and CN. The addition of $O_2$ to the $C_4F_7N/CO_2$ gas mixture can effectively reduce the decomposition of $C_4F_7N$ and the generation of most of the by-products. The decomposition of $C_4F_7N$ is the least when the $O_2$ ratio is 6%. The addition of 0%-4% $O_2$ decreases the rate of the main decomposition reactions in the reaction system, and the rate of the reaction is increased when the $O_2$ ratio is greater than 8%. However, the initial decomposition time of $C_4F_7N$ was significantly reduced and the generation rate of decomposed particles was accelerated when the simulated temperature was higher than 2600 K.

(2) The addition of $O_2$ complicates the dissociation process of $C_4F_7N$, and the dissociation process that occurs when $C_4F_7N$ combines with O is more likely to occur compared to the direct dissociation process of the $C_4F_7N$ molecule, which has a reaction enthalpy of -106.52 kcal/mol and an activation energy of 47.51 kcal/mol. The majority of the reactions between the O particles and the decomposed particles of $C_4F_7N$ belong to the exothermic process that can occur spontaneously. This may prevent the particles generated by $C_4F_7N$ decomposition from recombining to form $C_4F_7N$ or other stable by-products, thus affecting the decomposition of the $C_4F_7N$ gas mixture and the formation of by-products.

(3) According to the simulation results, the addition of 2%-6%$O_2$ can appropriately inhibit the decomposition of $C_4F_7N$ and the generation of by-products, so the addition of 2%-6%$O_2$ can improve the chemical stability of $C_4F_7N/CO_2$ gas mixture.

## Ⅴ.REFERENCE

[1] Fu Y, Yang A, Wang X, et al. "Theoretical study of the neutral decomposition of $SF_6$ in the presence of $H_2O$ and $O_2$ in discharges in power equipment". Journal of Physics D: Applied Physics, vol:49, no:38, pp:385203, Aug 2016, doi: 10.1088/0022-3727/49/38/385203.

[2] Rabie M, Franck C. "Assessment of eco-friendly gases for electrical insulation to replace the most potent industrial greenhouse gas $SF_6$". Environmental Science & Technology vol:52, no:2 pp:369-380, 2018, doi: /10.1021/acs.est.7b03465.

[3] Zhang B, et al. "Determination and assessment of a complete and self-consistent electron-neutral collision crosssection set for the $C_4F_7N$ molecule." Journal of Physics D: Applied Physics vol:56, no:13, pp: 134001,2023, doi: 10.1088/1361-6463/acbd5d.

[4] Meyer F, Kieffel Y. "Application of Fluorontrile/$CO_2$/$O_2$ mixtures in high voltage products to lower the environmental footprint CIGRE reports". Paris, France: CIGRE,2018.

[5] Lindner C, Gautsch D. "Application of a fluoronitrile gas in a 123kV GIS pilot substation. CIGRE reports". Paris, France: CIGRE,2018.

[6] Zhou R, Chen J, Xiao S, et al. "Compatibility and interaction mechanism between the $C_4F_7N/CO_2/O_2$ gas mixture and FKM and NBR". ACS OMEGA, vol:8, no:12, pp:11414-11424, 2023, doi:10.1021/acsomega.3c00195.

[7] Perez, Antoine, et al. "Measurement of streamer propagation velocity over solid insulator surface in a $C_4F_7N/CO_2/O_2$ mixture under lightning impulse voltages." IEEE Transactions on Dielectrics and Electrical Insulation, vol:28, no:2, pp: 481-487,2021, doi:10.1109/TDEI.2020.009257.

[8] Zhang B, Zhou R, Hao M, et al. “Application study of $C_4F_7N$ gas mixture in 40.5kV circuit breaker (II): Arc extinguishing performance experiment and post-arc decomposition characteristics”. Chinese Journal of Electrical Engineering, vol:42, no:24 pp:9147-9159,2022, doi: 10.13334/j.0258-8013.pcsee.212624.

[9] Tong D, Zhao Q, Cao R, et al. “Experimental study on Initial partial Discharge Voltage of $C_4F_7N/CO_2/O_2$ Gas mixture.” High Voltage Technology, vol:49, no:03, pp:1007-1014,2022, doi: 10.13336/j.1003-6520.hve.20220778.

[10] Yang Y, Gao K, Yuan S, et al. “Study on power frequency breakdown characteristics of $C_4F_7N/CO_2/O_2$ mixture under typical electric field”. Transactions of China Electrotechnical Society, vol:37, no:15, pp:3913-3922,2019, doi: 10.19595/j.cnki.1000-6753.tces.220169.

[11] Li Y, Zhang X, Ye F, et al. “Influence regularity of $O_2$ on dielectric and decomposition properties of $C_4F_7N$-$CO_2$-$O_2$ gas mixture for medium-voltage equipment”. High Voltage, vol: 5, no:3, pp: 256-263,2020, doi: 10.1049/hve.2019.0219

[12] Ye F, Zhang X, Li Y, et al. “Effect of $O_2$ on AC partial discharge and decomposition behavior of $C_4F_7N/CO_2/O_2$ gas mixture”. IEEE Transactions on Dielectrics and Electrical Insulation, vol:28, no:4, pp: 1440-1448,2021, doi: 10.1109/TDEI.2021.009626.

[13] Yang Y, Gao K, Bi J, et al. “Effect of micro-oxygen and atmospheric pressure on decomposition characteristics of $C_4F_7N/CO_2/O_2$ mixture under suspension discharge”. High Voltage Technology, vol:47, no:10, pp:3566-3580,2021, doi: 10.13336/j.1003-6520.hve.20211100.

[14] Ye F, Zhang X, Xie C, et al. “Effect of oxygen and temperature on thermal decomposition characteristics of $C_4F_7N$-$CO_2$-$O_2$ gas mixture for MV equipment”. IEEE Access, vol:8, no:12, pp: 221004-221012,2020, doi:10.1109/ACCESS.2020.3043334.

[15] Fu Y, Wang X, Wang X, et al. “Theoretical study on decomposition pathways and reaction rate constants of $C_4F_7N$ with O atom”. Journal of Physics D: Applied Physics, vol:53, no:10, pp: 105202,2019, doi: 10.1088/1361-6463/ab5739.

[16] Zhang X, Li Y, Chen, D, et al. “Reactive molecular dynamics study of the decomposition mechanism of the environmentally friendly insulating medium $C_3F_7CN$”. RSC advances, vol:7, no:80, pp: 50663-50671,2017, doi:10.1039/C7RA09959B.

[17] Liu Y, Hu J, Hou H, et al. “ReaxFF reactive force field development and application for molecular dynamics simulations of heptafluoroisobutyronitrile thermal decomposition”. Chemical Physics Letters.vol:751, pp:137554,2020, doi: 10.1016/j.cplett.2020.137554.

[18] Nomura K. I, Kalia R. K, Nakano A, et al. “A scalable parallel algorithm for large-scale reactive force-field molecular dynamics simulations”. Computer Physics Communications.Vol:178, no:2, pp: 73-87,2008, doi: 10.1016/j.cpc.2007.08.014.

[19] Brault, P., Abraham, M., Bensebaa, A., Aubry, O., Hong, D., Rabat, H., & Magureanu, M. “Insight into plasma degradation of paracetamol in water using a reactive molecular dynamics approach”. Journal of Applied Physics. Vol:129, no:18, doi: 10.1063/5.0043944

[20] Mueller J E, van Duin A C T, Goddard III W A. “Development and validation of ReaxFF reactive force field for hydrocarbon chemistry catalyzed by nickel”. The Journal of Physical Chemistry C, vol:114, no:11, pp: 4939-4949,2010, doi:10.1021/jp9035056.

[21] Perdew J P, Burke K, Ernzerhof M. “Generalized gradient approximation made simple”. Physical review letters, vol:77no:18 pp: 3865,1996, doi:10.1103/PhysRevLett.77.3865.

[22] Halgren T A, Lipscomb W N. “The synchronous-transit method for determining reaction pathways and locating molecular transition states”. Chemical Physics Letters, vol:49, no:2 pp: 225-232,1977, doi: 10.1016/0009-2614(77)80574-5.

[23] Fu Y, Yang A, Wang X, et al. “Theoretical study of the decomposition mechanism of $C_4F_7N$”. Journal of Physics D: Applied Physics, vol:52, no:24 pp: 245203,2019, doi: 10.1088/1361-6463/ab0de0.

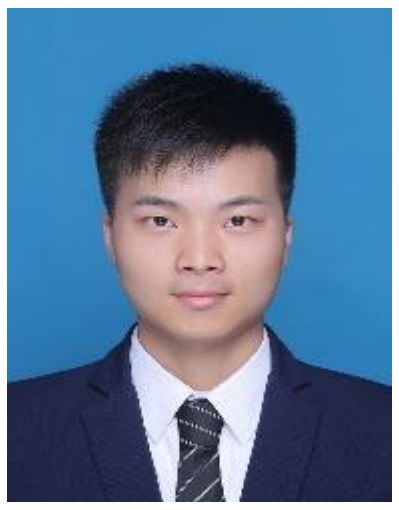

**Fanchao Ye** was born in Honghu City, Hubei Province, China, in 1993. He received his Doctor’s degree in electrical engineering from Wuhan University, Wuhan, China. Dr Ye is currently a lecturer with the School of Electrical and Electronic Engineering, Hubei University of Technology. His research interests include eco-friendly gas insulating media and fault diagnosis of high voltage electrical insulation equipment.

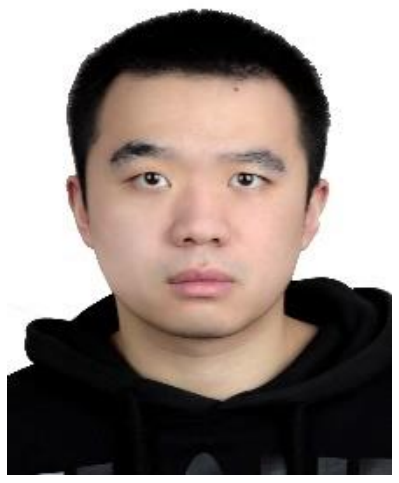

**Yitian Chu** was born in Honghu City, Hubei Province, China in 2000. He majored in Electrical Engineering at the School of Electrical and Electronic Engineering, Hubei University of Technology in 2023. He is mainly engaged in eco-friendly gas insulating media and fault diagnosis of high voltage electrical insulation equipment.

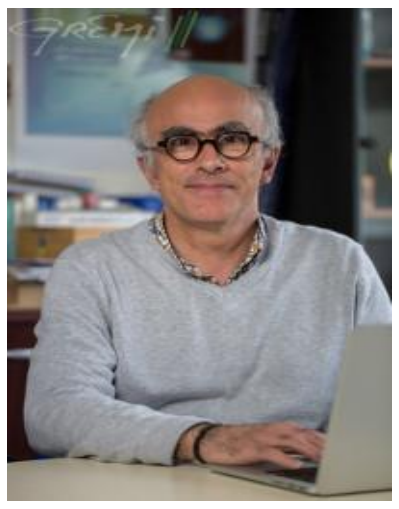

**Pascal Brault** is researcher at the French National Centre for Scientific Research. His research interests include modelling of plasma sputtering deposition, thin film growth and plasma chemistry using classical and ab-initio molecular dynamics as well as DFT/ quantum chemistry methods.

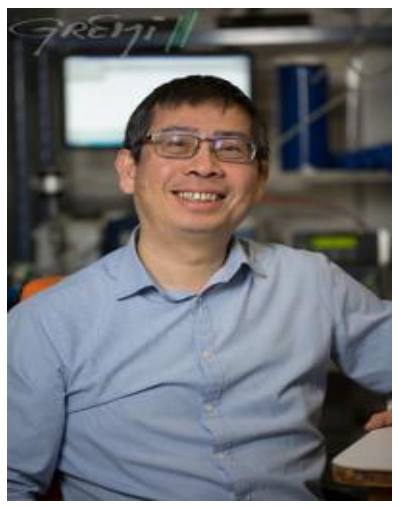

**Dunpin Hong** received his M.Sc degree from the University of Grenoble in France and his Ph. D degree in plasma spectroscopy from the University of Orléans in France. He is currently a Professor at the University of Orléans and works at the laboratory GREMI, a joint laboratory of C.N.R.S. and the University of Orléans. His research interests include plasma production by electric discharge, optical diagnostics of transient plasmas and non-thermal plasmas for subsonic airflow control and for water pollution control. Dr. Hong is a member of the French associations AAE (Association des Arcs Electriques) and EEA (Electronique, Electrotechnique, Automatique). He is a former IEEE member.

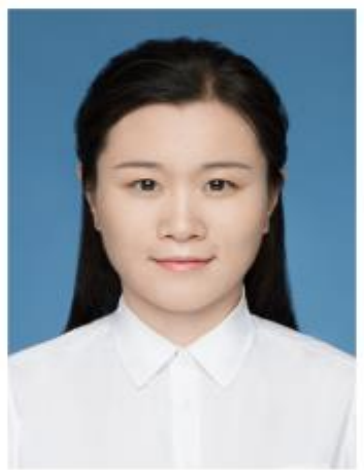

**Shuangshuang Tian** was born in Shandong, China in 1989. She is a Ph.D./postgraduate supervisor. She is mainly engaged in the online monitoring of power transmission and transformation equipment and application technology of eco-friendly gas insulating media.

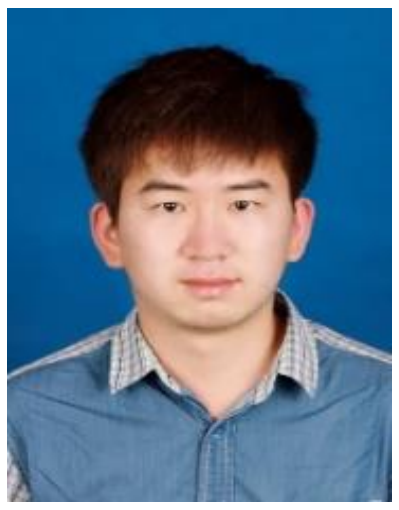

**Yi Li** was born in Shangluo City, Shanxi Province, China, in 1994. He received his Doctor’s degree in electrical engineering from Wuhan University, Wuhan, China. Dr. Li is currently an associate researcher at Wuhan University. His research interests include eco-friendly gas insulating media and fault diagnosis of high voltage electrical insulation equipment.

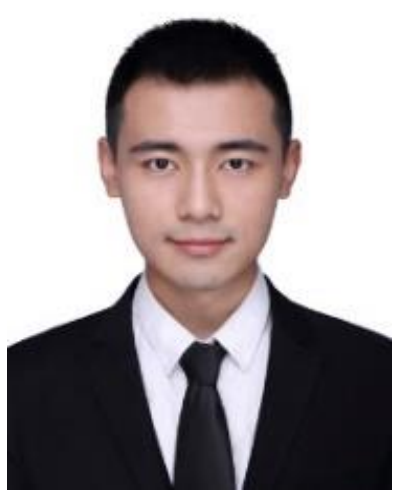

**Song Xiao** was born in Zhangjiakou City, Hebei Province, China, in 1988. He received his Ph.D. degree in electrical engineering at Chongqing University, Chongqing, China, and his Ph.D. degree in plasma engineering at Université de Toulouse, Toulouse, France. Dr. Xiao is an associate professor at the School of Electric Engineering, Wuhan University. His research interests include partial discharge online monitoring and eco-friendly gas insulating media.

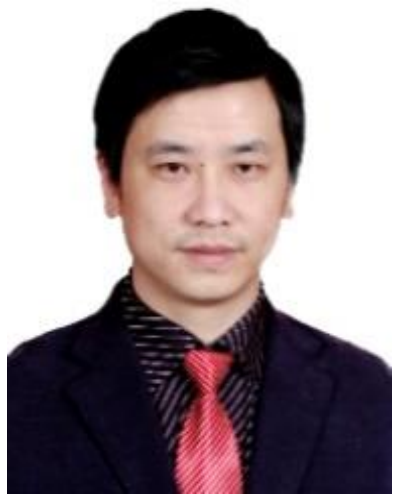

**Xiaoxing Zhang** was born in Qianjiang City, Hubei Province, China, in 1972. He received his Bachelor and Master degrees from Hubei University of Technology, and his Doctor’s degree from Chongqing University. Prof. Zhang is currently the dean of the School of Electrical and Electronic Engineering at Hubei University of Technology. His research interests include the online monitoring and fault diagnosis of high voltage electrical insulation equipment, eco-friendly gas insulating media, and new nano-sensors.